\documentclass[journal = jpcbfk, manuscript= article]{achemso}

\usepackage{amsmath}
\usepackage{multibbl}
\usepackage{amssymb}
\usepackage{textcomp}
\usepackage{ulem}
\usepackage{hyperref}
\usepackage{graphics}
\usepackage{graphicx}
\usepackage[export]{adjustbox}
\hypersetup{colorlinks,citecolor=black,filecolor=black, linkcolor=black, urlcolor=black}

\usepackage{caption}
\usepackage{float}
\usepackage{natbib}
\usepackage{setspace}
\usepackage{xkeyval}
\usepackage{ulem}

\usepackage[usenames,dvipsnames]{xcolor}
\usepackage{lipsum}

\newcommand{\MADa}{Departamento de F\'{i}sica Te\'{o}rica de la Materia
                   Condensada, Universidad Aut\'{o}noma de Madrid, E-28049 Madrid, Spain}
\newcommand{\MADb}{Department of Macromolecular Structures, Centro Nacional de 
                   Biotecnología, Consejo Superior de Investigaciones Científicas, 28049 
                   Cantoblanco, Madrid, Spain}
\newcommand{\MADc}{Condensed Matter Physics Center (IFIMAC),
                   Universidad Aut\'{o}noma de Madrid, E-28049 Madrid, Spain}
\newcommand{\basel}{Department of Physics, University of Basel, Klingelbergstrasse 82, 
                    4056 Basel, Switzerland}
\newcommand{\jena}{Otto Schott Institute of Materials Research, Friedrich Schiller 
                   University Jena, D-07742 Jena, Germany}
\author{R\'emy Pawlak}
\affiliation{\basel}
\email{remy.pawlak@unibas.ch}
\altaffiliation{Contributed equally to this work}
\author{J.G. Vilhena}
\affiliation{\MADa}
\affiliation{\MADb}
\alsoaffiliation{\basel}
\altaffiliation{Contributed equally to this work}
\affiliation{\basel}
\author{Antoine Hinaut}
\affiliation{\basel}
\author{Tobias Meier}
\affiliation{\basel}
\author{Thilo Glatzel}
\affiliation{\basel}
\author{Alexis Baratoff}
\affiliation{\basel}
\author{Enrico Gnecco}
\affiliation{\jena}
\author{Rub\'en P\'erez}
\affiliation{\MADa}
\altaffiliation{\MADc}
\email{ruben.perez@uam.es}
\author{Ernst Meyer}
\affiliation{\basel}
\email{ernst.meyer@unibas.ch}

\title{Conformations and cryo-force spectroscopy  of spray-deposited single-strand DNA on gold}
\begin{document}

\newpage
\noindent
\textbf{Cryo-electron microscopy has become a valuable tool to determine the structure of biological matter in vitrified liquids. So far, however,  mechanical properties of biomolecules, including elasticity and adhesion, have mainly been probed at room temperature using tens of pico-newton forces. Their detection is then limited by entropic fluctuations causing unfolding-refolding events. Here, we combine scanning probe microscopy, force spectroscopy and computer simulations in cryogenic conditions to quantify intra-molecular 
properties of spray--deposited single--strand DNA  oligomers on Au(111). Images with sub--nanometer resolution reveal their folding conformations further confirmed in detail by molecular dynamics simulations. Single-chain lifting shows a progressive decay of the measured stiffness with sharp dips every 0.2-0.3 nm associated with sequential peeling and detachment  of single nucleotides.  An intra-molecular stiffness of 30-35 N.m$^{-1}$ per stretched ssDNA repeat unit is obtained in the nano-newton range. 
}\\

Nucleic acids (NA)~\cite{Sinden1994} are among the most studied bio--molecules nowadays due to their biological relevance but also in applications to nano-devices or computing~\cite{Seeman2017,Rothemund2004,Seeling2006}. Control over nucleotide sequences as well as knowledge of their folding properties has enabled the rational design of highly elaborate two- and three-dimensional DNA structures, so-called ”DNA origami”~\cite{Han2011} programmed by Watson--Crick complementarity~\cite{Crick1953}. These remarkable advances have been made possible by the accurate determination of nucleotides characteristics beforehand,  using non-invasive  single-molecule manipulation techniques such as optical tweezers~\cite{Smith1996,Neuman2004,Bustamante2003} or magnetic tweezers~\cite{Strick1996,Strick2000}. Beside these approaches, force spectroscopy based on atomic force microscopy (AFM) also allows direct measurements of mechanical, adhesion~\cite{Sonnenberg2007}  and tribological properties~\cite{Kühner2006}, as well as visualizing self-assembly processes.  So far, such force spectroscopic experiments have been conducted under ambient conditions in solutions, mostly up to few tens of pico-Newton tensile loads. Mechanical properties are then dominated by thermal fluctuations and folding/unfolding of soft parts~\cite{Lee1994,Manohar2008,Duwez2011,Iliafar2014}. Only few AFM studies on long polymers strongly bound at both ends reached the nN force level where thermal fluctuations are largely suppressed~\cite{Hugel2005}. To our knowledge, no features attributable to sub-nanometer structural details have been observed in force versus extension curves recorded under ambient conditions. Here we demonstrate that dynamic AFM--based force-spectroscopy in cryogenic conditions is a promising method for characterizing mechanics of DNA down to the sub-nm level.  Similar to the advent of cryo-electron microscopy for structure characterization of bio--systems~\cite{Dubochet1985}, further investigations  along this line 
 could open new avenues towards the integration of DNA into solid nano-devices through bio--mechanical studies at to this level of precision.\\

\subsection*{Real--space imaging of spray--deposited ssDNA on gold}  
Imaging of DNA has been a long-term challenge. Although numerous groups successfully visualized DNA branches with impressive spatial resolutions in solution under ambient conditions~\cite{Hansma1995,Bustamante1996,Dietler2013,Ares2016}, the highest accuracy have been reached using STM/AFM imaging at cryogenic temperatures enabling reduced contamination~\cite{Clemmer1991,Zhang1996,Tanaka2009}. These experiments also required efficient deposition techniques to successfully transfer macromolecules from solution onto a substrate while maintaining UHV cleanliness standards~\cite{Fenn1989,Rauschenbach2006,Hinaut2017}. However, the 
characterization of adsorbed bio--molecules at the sub-nm level, specifically of DNA under such conditions~\cite{Rauschenbach2006}, still remains rather unexplored.\\
\begin{figure*}[t!]
\centering
\includegraphics[width=0.75\textwidth]{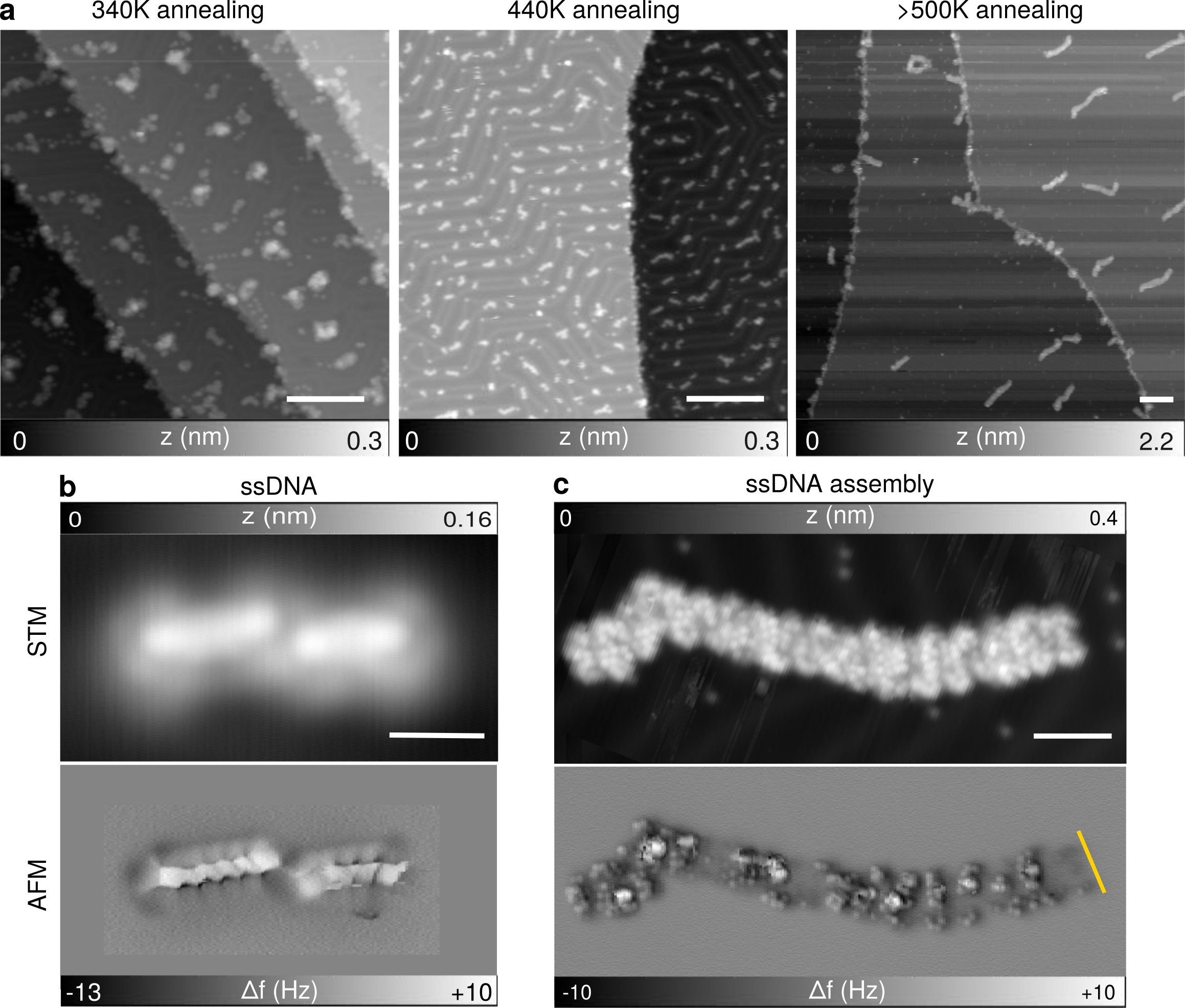}
\caption{{\bf Images of ssDNA morphologies as a function of Au(111) annealing temperature.} {\bf a,} STM overviews of spray-deposited ssDNA on Au(111) after annealing at 340~K, 440 K and above 500 K respectively (I$_t$ = 2 pA, V = - 1.3 V);  scale bars = 20 nm. {\bf b,} STM topographic image of a single 20-cytosine ssDNA oligomer after 440 K annealing (I$_t$ = 2 pA, V = - 1.3 V) and the corresponding higher resolution constant-height AFM image, both acquired with a CO-terminated tip; scale bar = 1 nm. {\bf c,} STM topographic image of self-assembled sDNA oligomers after 500 K annealing (I$_t$ = 2 pA, V = - 1.3 V) and the corresponding AFM image; scale bar = 4 nm. The yellow line highlights the length of a single ssDNA oligomer.}
\label{Fig1}
\end{figure*}

We exploit recent advances in frequency modulation AFM~\cite{Gissiebl2003}  under cryogenic conditions (see Methods), that pushed spatial resolution of adsorbed molecules to the single-bond level~\cite{Gross2009} and enabled complex molecule manipulations at surfaces~\cite{
Wagner2012,Kawai2014,Kawai2016,Pawlak2016b}, to demonstrate imaging and manipulation of single-strand DNA (ssDNA) 20-cytosine oligomers  at low temperature (5K). The oligomers were spray-deposited on the Au(111) kept in ultra-high vacuum (UHV) at room temperature (see Methods)
then annealed step by step up to a maximum temperature  $T_{max }$ = 500 K. 
In all cases the resulting structures were subsequently imaged at 5 K using constant-current STM. As shown in Fig.~\ref{Fig1}a, the adopted conformations evolve from clusters of $\sim$ 5-8 nm diameter to 4 nm long isolated oligomers upon annealing below 500 K. When $T_{max }$ $\geq$ 500 K, the oligomers coalesce into several nano-meter long structures. Sub-nm contrast could be obtained along individual oligomers not only by STM, but even better using constant-height AFM with CO-terminated tips~\cite{Gross2009}  (Fig.~\ref{Fig1}b and c). Note that 
water contamination becomes quite low upon annealing to 440K (Supp. Info Fig. S1).\\
\begin{figure*}[t!]
\centering
\includegraphics[width=\textwidth]{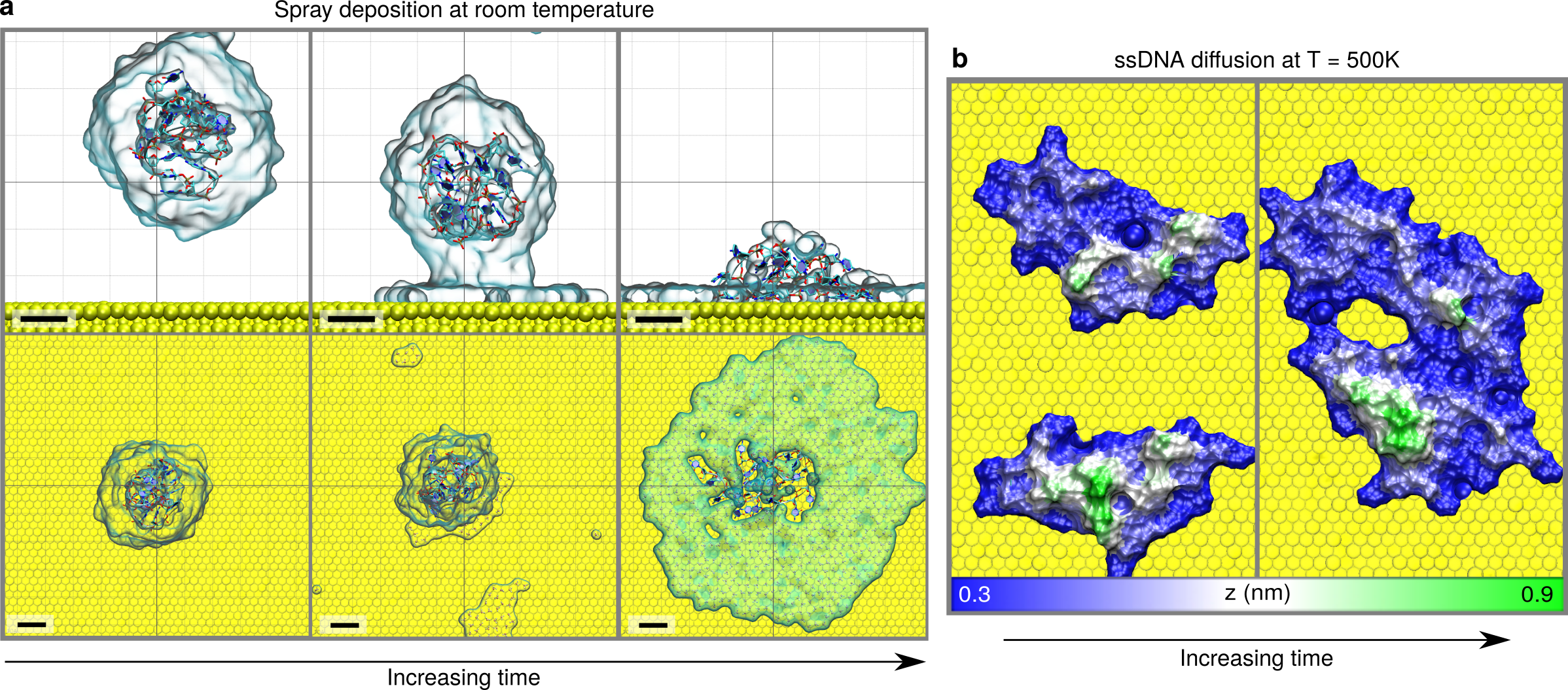}
\caption{{\bf Molecular dynamics simulations of spray--deposition and diffusion of ssDNA on Au(111).} {\bf a,} Side-  and top views of a small water droplet containing one ssDNA oligomer getting adsorbed on the gold surface. Water is represented using a transparent surface;  scale bars = 1 nm. {\bf b,} 500 ns--MD simulation of two oligomers assembled by diffusion at 500 K;  their initial configurations 
were obtained from a simulation of single ssDNA oligomer adsorption in vacuum  (Supp. Info Fig. S2d). At 500~K, both oligomers start diffusing, thus favouring self-assembly assisted by intermolecular interactions.}
\label{Fig2}
\end{figure*}

The previous observations are the result of enhanced ssDNA mobility at increasing temperatures, as we are able to confirm by MD simulations (see Methods).  Here one 20--cytosine ssDNA oligomer generated from the canonical B-form~\cite{Sinden1994} (Supp. Info. Fig. S2) was inserted into a water nano-droplet together with 19 charge-compensating Na$^+$ ions and virtually "sprayed" onto an unreconstructed Au(111) surface at room temperature (Fig.~\ref{Fig2}a).  Prior to deposition, the droplet surface tension causes a 
considerable folding of the ssDNA chain in the droplet into a compact structure only 3 nm in diameter (Supp. Info Fig. S4). Much less folding was observed if the chain was completely immersed in water, having a total length of 6.4 nm (Supp. Info. Fig. S3).  In the present case (Fig.~\ref{Fig2}a), a folded conformation is retained upon adsorption, but the structures became somewhat longer ($\sim$ 4 nm) compared to those embedded in the free droplet, in excellent agreement with the experimental images (Fig.~\ref{Fig1}b and Supp. Info. Fig. S1b). Different initial ssDNA configurations with and without water molecules have also been considered but always led to very similar final configurations (Supp. Info Fig. S5). The interaction with gold induces a systematic flattening of the ssDNA structure with most nucleotide bases $\pi$-stacked nearly parallel to the surface like for cytosine on Au(111).~\cite{Rosa2012}. The effect of annealing on ssDNA/Au(111) has been also reproduced by MD simulations. Starting and final configurations are shown in Fig.~\ref{Fig2}b and Supp. Info. Fig. S6. Coalescence of ssDNA oligomers is observed only at T = 500 K which is in agreement with the experimental data (Fig.~\ref{Fig1}). Top views show that the oligomers essentially preserve the folded structure after self-assembly.  Their alignment  are consistent with the appearance of constant-height AFM images in Fig.~\ref{Fig1}c.

\subsection*{ssDNA cryo--force spectroscopy }
\begin{figure*}[t!]
\centering
\includegraphics[width=\textwidth]{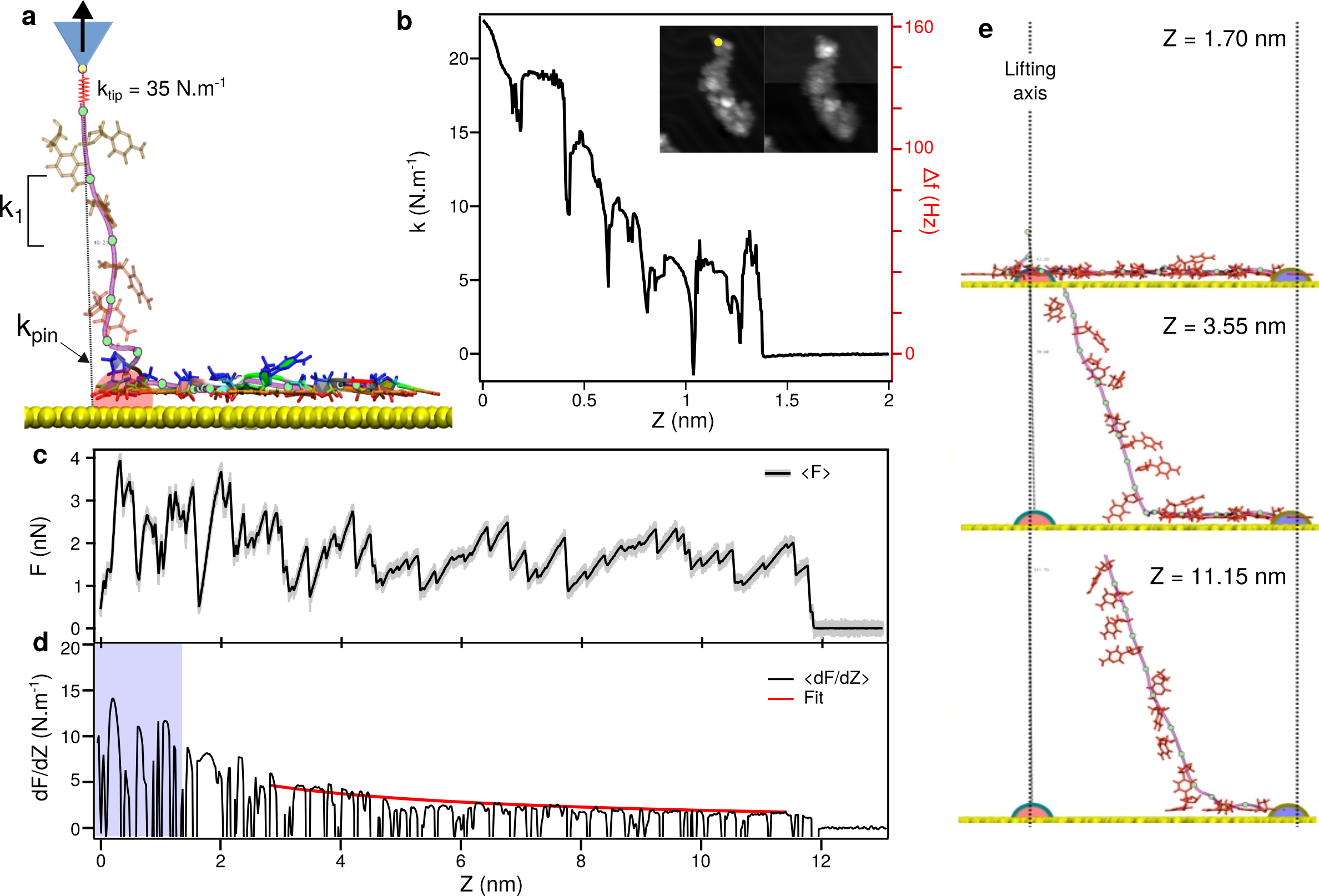}
\caption{{\bf Mechanical response while lifting a ssDNA oligomer from gold.} The oligomer is attached at one end to the AFM tip and pulled vertically at constant velocity. {\bf a,} Schematic of the lifting simulations. {\bf b,} Experimental retraction trace, $\Delta f(Z)$ $\propto$ $k(Z)$ recorded at 4.8 K. {\bf c,} Force-distance curve $F(Z)$ obtained from MD simulations assuming $k_{tip}$ = 35 N.m$^{-1}$ and {\bf d,} the corresponding computed stiffness $k(Z)$. The blue area shows the Z-range accessed by the experiment. The red curve shows the fit to $k$ maxima according to Eq. \eqref{1} with $k_1$ = 32.6 $\pm$ 3.9 N.m$^{-1}$, $k_{pin}$ = 18.3 $\pm$ 3.3 N.m$^{-1}$ and $b \cos \theta $ = 0.64 $\pm$ 0.05. 
{\bf e,} Side views of the ssDNA oligomer after lifting 0, 10, 17 nucleotides revealing the nearly straight configuration of the detached segment.
}
\label{Fig3}
\end{figure*}

To investigate the mechanical properties of ssDNA pre-adsorbed on Au(111), we have attempted to lift  single oligomers from the surface, either isolated ones or like those assembled see in Fig.~\ref{Fig1}b and c, respectively. We used the protocol introduced in Ref.~\cite{Grill2009} and used in Ref.~\cite{Kawai2014} to pull off single polyfluorene chains with the AFM tip. The tip apex was first gently indented into Au(111), then approached to the end of an oligomer (yellow dot in the left inset of Fig.~\ref{Fig3}b) until a contact was established, as revealed by a sudden increase in the $\Delta f$ and $I_t$ signals (Supp. Info Fig. S7). The zero value of the tip-surface distance Z was set  at that position, and the tip was then vertically retracted at constant speed $v$ = 22 pm.s$^{-1}$. Representative experimental and theoretical retraction traces are shown in Fig.~\ref{Fig3}. In Fig.~\ref{Fig3}b, the measured stiffness $k \approx 2k_0 \Delta f/f_0$  (see Methods),  $k_0$ being the deflection sensor stiffness and $f_0$ its resonance frequency, decreases progressively from 23 to about 5 N.m$^{-1}$ as the tip--sample separation $Z$ increases.  This variation is interrupted by  narrow dips observed every 0.2-0.3 nm followed by an abrupt drop to zero when the tip is approximately at 1.4 nm far from the contact point, well below the length of one ssDNA oligomer in its adsorbed configuration (4 nm) as seen in Fig.~\ref{Fig2}. The underlying premature detachment of the ssDNA from the tip (see Supp. Info S6 and S8) is confirmed by comparing STM images before and after the lifting experiment showing the oligomer still on the surface (Inset Fig.~\ref{Fig3}b).  Experimentally, only partial lifting of single ssDNA oligomers either self-assembled or individually adsorbed (Supp. Info Fig. S7) could be achieved.\\

Steered MD simulations~\cite{Park2003,Vilhena2016b} have been run to shed light on the experimental results. A virtual "tip atom" connected to the P atom of the backbone between the first two nucleotides by a spring of stiffness $k_{tip}$ representing the tip-molecule bond was pulled up at a constant speed of 0.1 m.s$^{-1}$ while keeping the substrate at a temperature of 5K (Fig.~\ref{Fig3}a). From the recorded variations of the normal force $F(Z)$ (Fig.~\ref{Fig3}c; see also Methods), the effective stiffness was extracted as $k$ = $dF(Z)/dZ$ (Fig.~\ref{Fig3}d). In the simulations the whole ssDNA oligomer detaches from the substrate when the tip has been retracted up to $Z_{off}$  = 11.8 nm, which is slightly less than its fully stretched length (see Supp. Info Fig. S8).  In spite of the discrepancy with the experimental value of $Z_{off}$, the measured maximum $k$ values exhibit a similar trend in the common Z-range (blue area in Fig.~\ref{Fig3}d). Not only the stiffness $k$ decreases from comparable initial values of $\sim$ 15-25 N.m$^{-1}$, but pronounced dips (coinciding with abrupt force drops in the simulations) also occur at repeat distances of about 0.2--0.25 nm, i.e. well-below the average base-base attachment distance $b$ along the stretched backbone 
(see Supp. Info Fig. S8). Careful observation of the configurations adopted by the ssDNA atoms during simulated pulling reveals that the observed repeat distances reflect intermediate stages (peeling, lifting , detachment) in the successive lifting of cytosine bases often 
accompanied by stick-slip-like sliding of the adjacent base over the Au(111) surface (0.28 nm lattice spacing) and irregular unfolding of the backbone.  Details can be visualized in the movies provided in the Supporting Information. The first steps of the lifting process involve correlated base detachments and intricate unfolding which require increased lifting forces. This increase might also cause the experimentally observed premature detachment of ssDNA from the tip apex.  
Similar observations have been reported at room temperature in solution for grafted polymers~\cite{Sonnenberg2007} and ssDNA adsorbed on carbon nano-tubes~\cite{Iliafar2014}.\\

It is worth noting that the computed $k$ variations as a function of $Z$ are remarkably different from those previously reported for polyfluorene chains lifted with the same method from Au(111)~\cite{Kawai2014},  where the $k$ maxima ($\approx$ 0.4 N.m$^{-1}$) were constant during retraction and the process ended at a distance corresponding to the number of monomers (initially identified by STM). In that case as well as graphene nanoribbons on the same surface~\cite{Kawai2016}, the much stiffer repeat units weakly adhere to the substrate, thus allowing nearly frictionless sliding followed by complete chain detachment. In the present case the ssDNA backbone is more flexible and most bases are strongly bonded to the gold surface. Snapshots from our MD simulation movies (Fig.~\ref{Fig3}e) indeed show that the bases remaining on the surface do not slide while the lifted segment between tip and sample becomes essentially straight but inclined by a constant angle $\theta$ between $Z \approx$ 5 and 11 nm. This underlying sequential base detachment, similar to peeling off an adhesive tape, reflects the strong adhesion of adsorbed cytosine bases. Interestingly, “infinite” as opposed to negligible friction was detected for long ssDNA chains in solution at room temperature when adsorbed on gold~\cite{Kühner2006} and graphite~\cite{Manohar2008}, respectively.  Easy sliding of short ssDNA oligomers on graphene under cryogenic conditions was recently predicted by steered MD simulations.~\cite{Vilhena2016b}\\

The gradual reduction of the $k$ maxima arises because the stiffness of the lifted segment decreases as it becomes longer. Focusing on the most pronounced $k$ maxima achieved on the longest nearly linear parts of $F(Z)$, we assume local mechanical equilibrium in the springs-in-series model applied earlier to polyfluorene chains and to unzipped dsDNA hairpins~\cite{Alemany2014}. Including the stiffnesses $k_{tip}$ and $k_{pin}$ of the segment ends anchored to the tip and to the  adsorbed part of the oligomer  (Fig.~\ref{Fig3}e and Supp. Info 8), the envelope of $k$ maxima is expected to satisfy: 
\begin{equation}
k = \Big[ \frac{1}{k_{ends}} + \frac{n}{k_1} \Big]^{-1}
\label{1}
\end{equation}
with: 
\begin{equation}
\frac{1}{k_{ends}} = \frac{1}{k_{tip}} + \frac{1}{k_{pin}},
\end{equation}
$k_1$ being the stiffness per repeat distance $b$ in fully stretched ssDNA and $n$ = int [$ Z/(b. cos~\theta)$] is the number of nucleotides detached from the substrate in the range where the lifted segments are straight and inclined by $\theta$ (almost 18$^{\circ}$ between n = 8 and 18, see Fig.~\ref{Fig3}e and Supp. Info movies). The resulting fit (red curve) is superimposed on the computed $dF/dZ$ trace in Fig.~\ref{Fig3}d obtained for $k_{tip}$ = 35 N.m$^{-1}$. The resulting parameters are $k_1$ = 32.6 $\pm$ 3.9 N.m$^{-1}$ and $k_{pin}$ = 18.3 $\pm$ 3.3 N.m$^{-1}$ which confirm that ssDNA is strongly adsorbed and that fully stretched ssDNA is much more compliant than polyfluorene~\cite{Kawai2014},  presumably because the stretched ssDNA backbone stiffness is dominated by bond angle bending.  As a consequence, $k_{DNA}$ = $k_1$/19 = 1.7 N.m$^{-1}$ would be the stiffness of the fully stretched 20-cytosine oligomer with one base at each end subject to an average tension of $\sim$ 2 nN. This load is one to two orders of magnitude larger than maximum values attained in typical room-temperature investigations of ssDNA~\cite{Lee1994,Sonnenberg2007,Manohar2008,Duwez2011,Iliafar2014,Alemany2014}. We obtained only slightly different results and fit parameters from independent simulations assuming higher and lower $k_{tip}$ -values (Supp. Info Fig. S8 and Fig. S9). In particular, $Z_{off}$, $k_1$, $b$ and the average tension in the fitting range did not change appreciably. 

\subsection*{Summary and Conclusions}
The present work relies on closely matched scanning tunneling and force measurements, and computer simulations. The first part addresses the adsorption and self--assembly of single--strand DNA cytosine oligomers spray--deposited on Au(111) at the sub-nanometer level.  In the second part, the mechanical response of single ssDNA oligomers lifted lifting from the gold surface is investigated at the same level during using cryogenic force spectroscopy.  Multi-stage detachment is inferred, similar to peeling off an adhesive tape. This behaviour reflects the strong adhesion of adsorbed ssDNA bases on gold which is also manifested in the high measured initial effective stiffness of $\sim$ 15 N.m$^{-1}$, as well as in the comparable pinning stiffness  extracted from molecular dynamics simulations and intrinsic stiffness per repeat unit of fully stretched ssDNA ($\sim$ 33 N.m$^{-1}$). Compared to  $k_{pin} \approx$ 0.7 N/m obtained for polyfluorene, the present estimate suggests that complete detachment of ssDNA might be achieved by using end linkers forming stronger bonds to gold in UHV.  Although more difficult than solution chemistry, this task appears feasible in view of the successful detachment of PTDCA from Au(111) following contact to a carboxylic oxygen atom in spite of the high adsorption energy~\cite{Wagner2012}.  Our results further suggest that cryogenic force spectroscopy based on  dynamic force microscopy has the potential to study strongly adsorbed biomolecules or similar nano-sized synthetic systems with sub-nanometer resolution under tensile loads up to a few nano-Newtons, ten to hundred times higher than hitherto applied in most single--molecule force spectroscopy studies under ambient conditions.


\section*{Acknowledgement}
A.B., A.H., E.M. and R.P, T.G and T.M. thank the Swiss National Science Foundation (SNF) and the Swiss Nanoscience Institute (SNI). The COST Action MP1303 is gratefully acknowledged.
R.P and JGV thank the financial support of the Spanish MINECO (projects MAT2014-54484-P and MAT2017-83273-R) and also the computer resources, technical expertise and assistance provided by the Red Espa{\~n}ola de Supercomputaci{\'o}n at the Minotauro Supercomputer (BSC, Barcelona). J.G.V. acknowledges funding from a Marie Sklodowska-Curie Fellowship within the Horizons 2020 framework.
Dr. Marcin Kisiel is acknowledged for fruitful discussion. 

\section*{Author contribution}
E.M., R.P., R.P. and J.G.V. conceived the experiments. A.H. performed the electrospray deposition. R.P. performed the STM/AFM measurements and lifting experiments. J.G.V. conducted the numerical calculations. R.P. and J.G.V. analysed the data and co-wrote the manuscript with the help of E.G. and A.B. All authors discussed on the results and revised the manuscript.

\section*{Competing interests.}
The authors declare no competing financial interests.

\section*{Additional information}
\paragraph*{Supplementary information} is available for this paper at !!!!. 

\paragraph{Reprints and permissions information} is available at www.nature.com/reprints. 

\paragraph{Correspondence and requests for materials} should be addressed to E.M., R.P. or J.G.V. 

\section{Methods}
\paragraph*{Sample preparation and electro--spray deposition.}
An Au(111) single crystal purchased from Mateck GmbH was cleaned by several sputtering and annealing cycles in a ultra-high vacuum (UHV). The single-strand DNA molecules, purchased from Microsynth, were further diluted to dissolved with a concentration of around 1 nmol.ml$^{-1}$ in a mixture of water and methanol (ratio of 4:1) and then sprayed in UHV (see Supp. Info). Depending on the time varies from 1 to 30 min at constant solution flux controlled by a syringe pump with a speed of 2-10 $\times$ 10$^{-6}$ l.min$^{-1}$. Further details on the electro-spray-deposition apparatus and characterization can be found in Refs.~\cite{molecularspray,Satterley2007,Hinaut2015}.

\paragraph*{STM/AFM microscopy.}
The STM/AFM experiments were carried out at $\sim$ 5 K with an Omicron GmbH low-temperature STM/AFM controlled by a Nanonis RC5 electronics. We used commercial tuning fork sensors in the qPlus configuration~\cite{Gissiebl2003bis}, e.g. one prong fixed, the other with an etched tungsten tip epoxied at the end (unperturbed frequency  $f_0$ = 26 kHz, quality factor $Q$ = 10000-25000 in UHV, nominal spring constant $k$ = 1800 N.m$^{-1}$. These tips were sharpened by pre-indention into gold; some were then terminated by a CO molecule at the apex picked up from the surface. All voltages refer to the sample bias with respect to the tip. The constant-height AFM images were acquired with CO-terminated tips using the non-contact mode by driving the free prong on resonance while maintaining a constant tip oscillation amplitude $A$ = 50 pm. . 

\paragraph*{Lifting experiments.}
Pulling experiments were performed under the same conditions with gold--decorated tips while simultaneously recording the tunneling current at a typical bias of 40 $\mu$V. The ssDNA oligomers were picked up by gently pressing the AFM tip to the molecule at one of its extremities. Attachment of the molecule to the apex is revealed by an abrupt jump in the force and current signals as shown in Supp. Info. Fig. S7. Force spectroscopic measurements upon retraction were performed at a velocity of 22~pm.s$^{-1}$. In contrast to such measurements on biomolecules in ambient conditions, the gradient of the force along the oscillation direction rather than the pulling force itself is thus measured here.

\paragraph*{Atomic Level Models and Force Fields.}
In our simulation we consider one Au(111) surface of a three atomic layers thick slab, where the positions of the atoms in the lowest layer are fixed during the MD runs using a harmonic restrain of 5~Kcal.mol$^{-1}$. We considered surfaces of two different sizes, i.e. 16$\times$16~nm$^2$ (Fig.~\ref{Fig2} and Fig. S5) and 18$\times$22~nm$^2$ (Fig.~S6). The initial structure  was generated using the software NAB~\cite{NAB} thus obtaining a double helix with the canonical B-form as shown in Fig. S2. Then we removed the complementary sequence and used only the single-stranded 20--cytosine mer. Given that the phosphate groups in the backbone of the ssDNA are charged we added 19 sodium counter--ions. 
The ssDNA atoms were described using both the parmbsc0~\cite{parm} and the $\chi$OL3 refinements~\cite{chi} of the Cornell ff99 force field~\cite{ff99}. The choice of this force field is motivated by the fact that it is able to accurately describe the mechanical properties of DNA~\cite{Vilhena2017_DNAFlex} as well as adsorption of biomolecules to surfaces~\cite{Vilhena2016_IgG}.
The sodium counter-ions were described using the recently improved Joung/Cheatham parameters~\cite{Joung1,Joung2}.
As for the gold atoms, we resorted to CHARMM-METAL force-field~\cite{Heinz2013,HeinzREV} which simultaneously describe the intrinsic properties of gold, while retaining thermodynamical consistency with all the other force fields used here~\cite{Heinz2013,HeinzREV}. This force field has been extensively tested by studying the adsorption of different peptides (charged and uncharged) against both density-functional-theory simulations as well as available experimental results\cite{Heinz2013,HeinzREV}. In the simulations performed in water (Fig.~\ref{Fig2} and Sup. Info. Fig. S5 ), the water molecules are explicitly modeled using the TIP3P force field~\cite{TIP3P}.

\paragraph*{Molecular Dynamic (MD) Simulations.}
MD simulations were carried out using AMBER14 software suite~\cite{NAB} with NVIDIA GPU acceleration~\cite{nvidia1,nvidia2,nvidia3}. 
Periodic boundary conditions and Particle Mesh Ewald (with standard defaults and a real-space cutoff of 2 nm) were used to account for long-range electrostatic interactions. Van der Waals interactions were truncated at the real space cutoff, and Lorentz-Berthelot mixing rules were used to determine the interaction parameters between different atoms. In all vacuum simulations the volume of the system was kept fixed and the temperature was adjusted by means of a Langevin thermostat with a damping rate of 1~ps$^{-1}$ ensuring fast thermalization. The SHAKE algorithm was used to constrain bonds containing hydrogen, thus allowing us to use an integration time step of 2 fs. Coordinates were saved every 1000 steps. In all our simulations we observed that the final configuration was stable as it did not change during the last 40~ns of simulations (which was corroborated by the low , i.e. $<$ 0.2 nm , root-mean-square deviation). 
In the steered-molecular-dynamics simulations illustrated in Fig. 3 and Fig. S8 a harmonic potential of strength $k_{tip}$ moving with the tip was added and the force $F(Z)$ was computed directly as $k_tip (Z -\left<Z_P\right>)$ ~\cite{Kawai14,Kawai16,Vilhena18} which is justified at temperatures such that refolding or rebonding are not observed. Otherwise $F(Z)$  is the Z-derivative of the free energy including such rare but observable events, evaluated like in the method proposed by Park {\it et al}~\cite{Park2003bis}.

 \begin{tocentry}
 \newpage
 \appendix
 \centering
\includegraphics[height=3.55cm]{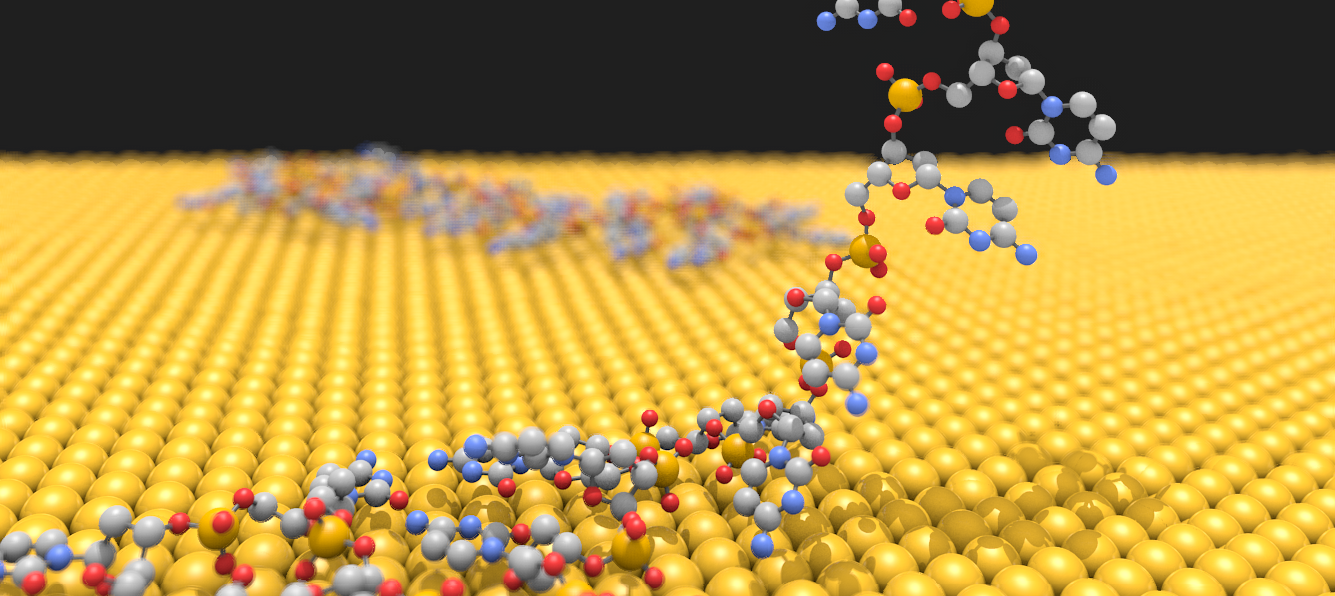}\\
\bigskip
 \end{tocentry}

\end{document}